\documentclass[a4paper]{article}
\usepackage{RR}
\usepackage{hyperref}
\usepackage{graphicx}
\usepackage{amsmath}
\usepackage{amsfonts}
\usepackage{amssymb}
\usepackage{fill}
\usepackage{makeidx}
\usepackage{url}
\usepackage{color}
\usepackage{latexsym}
\usepackage{anysize}
\RRdate{Juin 2006}
\RRauthor{Samuel Galice 
\and 
V\'eronique Legrand 
\and Marine Minier 
\and John Mullins\thanks{The professor John Mullins from the Ecole Polytechnique de Montr\'eal join the research while visiting the CITI Laboratory and the INRIA ARES project during his sabbatical}
\and St\'ephane Ub\'eda}

\authorhead{Galice \& Legrand \& Minier \& Mullins \& Ub\'eda} 
\RRtitle{Le projet KAA : un point de vue sur la politique de confiance} 
\RRetitle{The KAA project: a trust policy point of view} 
\titlehead{A trust policy} 

\RRresume{Dans les réseaux \textit{ambiants} où chaque entité doit faire confiance à ses voisins plutôt qu'à un réseau fixe, nous proposons dans ce rapport un modèle de gestion de confiance fondé sur notre vision des modèles sociaux. Chaque noeud construit ici un niveau local de confiance ne tenant pas compte de recommandations que pourraient fournir les autres noeuds. Ce paramètre dépend uniquement du nombre de noeuds de confiance rencontrés en commun avec le noeud avec lequel il veut intéragir. Chaque rencontre avec un noeud de confiance permet de créer un élément dans une base de donnée locale appelée \textit{historique}. Afin de garantir la sécurité des éléments de cet historique et leur non-transférabilité, nous utilisons un protocole cryptographique dédié à cette utilisation appelé CHE pour Common History Extraction.
}
\RRabstract{In the context of \textit{ambient networks} where each small device must trust
its neighborhood rather than a fixed network, we propose in this paper a \textit{trust management framework} inspired by known social patterns and based on the following statements: each mobile constructs itself a local level of trust what means that it does not accept \textit{recommendation} by other peers, and the only relevant parameter, beyond some special cases discussed later, to evaluate the level of trust is the number of common trusted mobiles. These trusted mobiles are considered as entries in a local database called \textit{history} for each device and we use identity-based cryptography to ensure strong security: history must be a non-tansferable object.
}
\RRmotcle{gestion de la confiance, réseau ad hoc, modèle de réputation local, protocole cryptographique}
\RRkeyword{trust management, ambient network, local reputation model, cryptographic protocol}
\RRprojet{ARES} 

\RRtheme{\THCom} 
 \URRhoneAlpes 
\begin{document}
\makeRR   

\tableofcontents
\section*{Introduction}
\addcontentsline{toc}{section}{Introduction}
Nowadays, wireless communications form an important aspect of smart and autonomous devices. They provide mobility and autonomous actions by offering open solutions for anywhere and anytime connectivity. In this way, the mobile as a node in a Personal Area Network is only a device in an environment where every object is able to communicate. In such a context, the capabilities offered by the wireless communications to smart objects will not be restricted simply to fixed networks accesses; rather, their peer-to-peer communication capabilities will have to receive more attention.

Devices in radio range can potentially establish self-organized networks of two or more objects. Mobility during the use of more complex services, is becoming commonplace and is addressed by means of ad hoc communication capabilities. However, mobile entities often become disconnected from their home networks and have to handle unforeseen circumstances. This notion of ambient networks also significantly complicates the deployment of services including the security. The device needs to carry self-contained informations and methods to be able to make fully autonomous security decisions. In such an environment where anonymous entities could build or exchange services, security and risk management could be seen as desired properties in order to ensure availability and robustness of these networks. Indeed, mobile may face with many threats as leaking of secret information, message contamination and node interposition, otherwise known as man-in-the-middle attacks.

The lack of a fixed network and the self-organized aspect are strong restricting constraints and a central trusted entity at each moment is inherently impossible. Moreover, active compromising of nodes is a real threat that requires strong security mechanisms. In a decentralized management system, security must fulfill several objectives. First, it must protect sensitive information against unauthorized access. Second, it must protect the distributed configuration algorithm from attacks. Third, it must prevent management functionality to be used as an attack tool, e.g., for flooding attacks.

Solutions exist as encrypted traffic which protects sensitive information; digital signatures allow verification of the authenticity of management communication, protecting the operation of the distributed algorithm and consequently mitigate the use of management functions for attacks. But these solutions drasticaly limit inter-operability between different security domains.

Thus, contrasting with strong security, \textit{trust management framework} is also needed in such an environment. Trust management allows us to use trust to achieve security goals. It is a larger framework that includes both risk management and access control. Its purpose is to mitigate the risks that are present in interactions. According to \cite{blaze96decentralized}, trust management is the capture, evaluation and enforcement of trusting intentions. It provides a unified approach to specifying and interpreting security policies, credentials, relationship which allows direct authorization of security-critical actions. In practice, a trust management scheme is usually based on a decision-making process coupled with a reputation and/or recommendation mechanism: neighbors of a smart device become the only source of informations. In general, mostly trust management systems consider trust as a threshold probability.

Trust management can quickly become too complex and tedious for people to understand and it is a necessity to have smart-trust systems which can reason about new situations automatically. Automated trust management allows us to trust software to make decisions about trust as we ourselves would do it. It allows existing devices to be trusted in news situations and by using evidence from trusted referres and past experience, it automatically forms new trusting beliefs.

If trust assumes some acceptance of risk, we need to minimise risk and maximise trust.
Unlike trust, risk management is well understood (Banking, Finance and Insurance). Trust increases efficiency in commerce or market. Trust is critical for e-commerce to remain dynamic and efficient. However, increased trust involves increased risk which must be managed.   Risks can be managed by using trusted third parties. We need to be able to capture trust from the natural environment. Risk management allows us combine risk with trust to form trusting intentions i.e. the security policy.

Access control uses authorisation information to determine whether access to resources should be allowed. This is similar to enforcing trusted intentions.  Access control policy enforces trusting behaviour.\\

This paper is organized as follows: section \ref{KAA} describes the context of the KAA project and how we use our social vision of the problem and how the trust is inherently present in the human society at each level. Section \ref{protocol} presents the cryptographic protocol carried by each node that permits  the trust bond to be proved or created and details some experimental results required for our protocol. Section \ref{reputation} presents the semantic notions inherently presented in our approach, the reputation model used and treats how to implement some social patterns into our model. Section \ref{further} releases the elements of a trust policy deduced from the semantic description of our model and describes the formal proof that will have to be done. Section \ref{conclusion} concludes this paper. 



\section{Context: the KAA project} \label{KAA}
The \textit{Knowledge Authentication Ambient project} (KAA project) is a collaborative research project involving research teams in computer sciences, mathematical modelling and social sciences. It proposes to look for human society trust management mechanisms and to derive an adapted technological version. We intend to develop a trust management framework including security protocols for devices that have no a priori relationship. Such a model will lead naturally to a decentralized approach that can tolerate some partial informations albeit there is an inherent element of risk for the trustee entity.

There is no one simple definition of trust. Instead, there is a system of definitions depending on which level of the problem we are looking at. Trust mechanisms are used by humans daily to promote interaction and they accept risk in situations where they have only partial informations. Trust allows us to reasonnably rely on the informations or actions of another party but it is an intrinsic and subjective property which may be propagated but not transferred. It is also a non-transitive notion: $A\rightarrow B\rightarrow C$ does not mean that $A\rightarrow C$. Trust is inherently linked to risk and can be viewed as a mechanism that permits to accept a certain risk when one interacts with others.

In this paper, we object the idea of a recommendation mechanism, trust is considered in our model in a non-transitive way. We opt for this limitation in order to avoid undesirable edge effets: rumour, bad opinions, etc. The principle of our approach is to base our decision-making process only on the actions which were beneficials. Consequently, a node could not receive a recommendation from another node that it has never met before because there is no simple local means to prove the authenticity of this recommendation. However, this approach permits a local reputation mechanism since it is based on certified events. 
\subsection{Social Patterns}\label{social}
Exchange is a central and traditional object within the social sciences, notably in economy where {\textquoteleft}market exchange{\textquoteright}  analyzes circulation of goods and services between agents (exchange is trust-regulated, that is to say sharing something enable unknown individuals to exchange), thus in sociology and in anthropology where the key concept is {\textquoteleft}social exchange{\textquoteright}, which gathers all kinds of non-economic exchange between individuals. Social patterns may be distinguish themselves on two strongly differentiating variables.  In one hand, the social distance that separates two individuals that exchange, this social distance being able to be strong in the case of the market and in the organization (this is the reason why the contract - commercial or labor - is so important to support exchange between unknowns) or, at the opposite, as often in the case of the family (included friends, neighbors, and other kind of strong social bonds and where exchange is gift-regulated) and network (as a community of individuals that share something like a life experience, an interest in something,\ldots) where familiarity, real or virtual, pre-exists to the exchange, allow individuals to exchange without contract and where exchange is trust-regulated.

On the other hand, the degree of structure of the institution, that defines the degrees of liberty of which the actors can dispose in order to exchange (notably the choice of the exchange partner and the choice of the exchanged things); this degree can be weak, as in the network and the market where individuals have all latitude to choose themselves and to exchange what they want to, or strong as in the family and the organization, institutions where exchange is more constrained by formal hierarchies and rules. Within each of these four patterns, exchange would be regulated by a major mechanism and a social distance:

\vskip5mm
\begin{center}
\begin{tabular}{cccc}
\hline
& & \multicolumn{2}{c}{\textbf{Social Distance}} \\
& & \textit{Strong} & \textit{Weak} \\
\hline	
\textbf{Degree of} & \textit{Strong} & Organization/Authority & Family/Gift \\ 
\textbf{structuration} & \textit{Weak} & Market/Price & Network/Trust \\
\hline
\end{tabular}
\end{center}
\vskip5mm

We think it is reasonable to construct a trust policy based upon social patterns that depends on the type of exchange. We can define two particular modes of operation: a closed mode where, as in the family case, the trust pre-exists and an open mode where, as in an organization (company), the trust needs to be established using a history-based protocol. In the closed mode, cryptographic verifications will not be absolutly required and in the open mode, trust policies will manage the degree of risk incurred. In short:
\begin{itemize}
\item \textbf{Family:} a community with a strong social distance and a strong degree of structure works in the closed mode.
\item \textbf{Network:} a community with a strong social distance and a weak degree of structure works in the open mode using a weak authentication for the nodes meetings.

\item \textbf{Market:} To implement a Market-like community, a money (either virtual or not) is required. Therefore, it is possible to add a semantic value to trustor proof and reciprocal proof based on the notion of market. If you add the value of the transaction in both proofs, you can achieve this goal. The trustor proof and reciprocal proof can be automatically generated when the transaction took place (security of this transaction is out of the scope of this paper).

\item \textbf{Organization:} In an organization pattern as a company, the closed mode is used with a strong authentication delivered by the imprinting station through classical certification. We could also integrate the hierarchical parameter in our model: for example, if someone wants to talk to a leader, the required number of common history elements need to be larger than if he wants to talk to a person at the same level than him.
\end{itemize}


We can also notice that depending on the considered social pattern, the semantic values associated to each history elements may be different. The process to derive the trust level from the size of the history, may also be rather different. We can consider that in a family-like pattern, only one interaction with another device of the group may be sufficient to accept interaction, whereas in a very big organization, the number of common elements between devices may graduate the authorization to various level of services.

\subsection{History based trust management framework} \label{framework}
History-based trust management has been yet studied in the past. In \cite{BussardMR04} such a trust management is presented but it is dedicated to a group signature and using trusty environment to generate elements of history. History-based trust management may be regarded as an consisting alternative proposal to a pairing model requiring frequently intervention of users. Pairing model is primarily relevant in the case of long term association between devices. For example, this approach is deployed in the built-in security mechanism of the Bluetooth chip. In this mechanism, the same identification information (a PIN code for example) has to be physically entered on each device. After the generation of a symmetric link key, devices are able to authenticating each other and to encrypt communication. The share PIN is the weakness of the model since it is prone to simple off-line attack \cite{WeakBluetooth}.
In \cite{SASN_Thomson}, a complete trust management framework is also proposed although it is only applied to a strongly closed environment where trust is a boolean value: the principle relies on removal or banishment of devices.

In our proposal, some basic informations are recorded in a history of past interactions between each mobile with an aim of reducing the incertitude on their behavior. Two nodes could exchange some services if they could trust one to each other and if they have a sufficient number of common previous interactions with some other nodes. We also specify our trust policy in terms of a semantic: A is trusted for B if condition X is true. We also have to provide security policy from our trusting beliefs and the possible threats by creating trusting intentions. Each elements of history are seen as a trust bond and are built upon cryptographic tools \cite{755/LCA}, \cite{yu00social}.

We aim to reduce the complexity of the decision-making process inherent of all trust management framework by relying the decision on the own-experience of each node. We use social pattern as a general framework to design our protocol. This framework includes many features such as emergence of spontaneous communities. But it is necessary in this case, to make some extra verifications in order to replace a trusted third party. Indeed, certify identity is a heavy shortcoming although the cryptographic foundations of our model is strong.

\subsection{The proposed model and security requirements}

All smart devices participating to this trust management architecture have to carry common specific cryptographic algorithms and protocols. This content is obtained through an \textit{imprinting phase} before any other interactions. Special fixed functional (supposed to be secure) units called \textit{imprinting stations} are supposed to exist \cite{SA99}. A device is belonging to a domain associated to a specific station and receives from these station an \textit{initial seed of trust} constructed from a master key $s$ unique for each station and kept secret. 

After this first interaction, each device's history is empty. Then, it starts recording an history based upon the knowledge of its interactions with some other nodes. The cryptographic protocol described in this paper ensures that any recorded element of history (generated for a specific encounter node involving two specific nodes) cannot be used by other nodes. This proved history is used to create and enforce the trust relation.

Before accepting any service interaction, two nodes must prove one to each other that they can establish a trust relation.  When two nodes meet for the first time, they search the common nodes they have met before. To prove those previous interactions, they have created together a provable value: they have respectively signed this value. Those common values also prove the identities of the nodes in presence signed by some other nodes.
The core method of the cryptographic method to do so is called \textit{Common History Extraction} protocol.
 
This method could be compared to the one used in the ``Web of trust'' defined by GnuPG \cite{PGP} done in a non transitive way. In our model, we prove an identity and previous interactions based on successful previous meetings but only at distance one between nodes. The identity itself is proved also by the use of identity-based cryptography and moreover we use elliptic curves cryptography as basic blocks to design our protocol.

It appears completely obvious that it is necessary to build at the same time the trust policies and the behavior of the system when taking account of the issue of risk assessment. Indeed, according to the number of common nodes and the nature of these elements in the respective histories, the fulfillment of a service might be conditioned by a certain number of criteria. For example, it is less risky to share its bandwidth with other nodes than offering a read (or write) access to confidential files. In  order to have a coherent model, the various issues involving in risk assessment must be studied.

Semantic value of each element of history plays a crucial role in our scheme. It permits to derive an information richer and more accurate than only the intersection of histories. This makes possible to establish an initial level of trust based not only on the \textit{quantity} of interactions, but also on its \textit{quality}. Such an extension is also out of the scope of this paper.

A trust notation or reputation principles as proposed for example in \cite{DBLP:conf/itrust/LiuI04} may be also added to the global architecture. But in our proposal, we prevent a subgroup of nodes from easily destroying a reputation of a particular node due to the fact that each result of interaction is cryptographically proved and thus, information upon the common nodes is not transitive. Compare to the PGP trust ring, an intermediate corrupted node could not destroy a complete trust chain relation due to the non transitivity of our trust establishment.

\subsubsection{The initial seed of trust.}


Each device receives a \textit{trust germ} from the \textit{imprinting stations}. It is composed by the following initial informations: $\mathrm{ID}$ an identity (eMail adress or IP address or just a simple name), $(S_{\mathrm{ID}},Q_{\mathrm{ID}})$ a first pair of secret/public key for cipher operations, a second pair of keys $(S_{\mathrm{ID}}^S,Q_{\mathrm{ID}}^S)$ for the signature and a set that represents all the public parameters of the elliptic curves required for the different computations. Those parameters are supposed to be the same for all the imprinting stations but none of those stations is supposed to be certified by any authority. Moreover, an independent mobile may be its own standalone security domain (it imprints itself). Another important point is that each smart device is sharing the same cryptographic algorithms and protocols downloaded from the imprinting station: a fingerprint algorithm, a signature algorithm, a zero-knowledge protocol, an algorithm to construct secure channel and the public parameters. The only values that each smart device has to keep  secret are $S_{\mathrm{ID}}$ and $S_{\mathrm{ID}}^S$ as usually in cryptosystem. 

In the context of mobile objects with low capacity, cryptography based on elliptic curves (ECC) leads to many advantages. In particular, the use of such cryptography makes it possible to develop algorithms and protocols of which the robustness and the cost in term of computation and space of storage are more advantageous than in usual cryptography.

\subsubsection{How the reciprocal trust may be enhanced.}

At the beginning of devices' life, just after each node has received its initial trust germ, history is obviously empty of any interaction. The number of common nodes is of course insufficient to permit an autonomous running and thus, it is necessary a bootstrap phase. So, two persons that want to exchange some services or some informations initiate an interaction by forcing by the hand this particular meeting - as in a Bluetooth like model, this gives the desired history element. After this bootstrap period, the nodes use the content of their histories to accept or reject a new interaction and the human intervention is then obvious. The benefit of our protocol takes all its dimension after this initial step very easy to use.

Then when two stranger nodes meet for the first time, they exchange their history and search their common nodes (the common nodes that they have met before). The interaction takes place if the number of common nodes is upper a given threshold. This kind of interaction is built upon the social principle ``a friend of my friend is a friend of mine''. Of course, they need to prove one to each other the common nodes that could be trusted. A first node Alice has already had a past interaction with Charlie, she may prove it to Bob if Bob also already had an interaction with the same Charlie. After a successful interaction, node Alice (resp. Bob) gives a proof of interaction to Bob (resp. Alice) that can be checked by any node having a past interaction with Alice (resp. Bob) in its history.

Our protocol also works for two nodes from stranger domains: the authentication notion is local and does not need to be strong. It comes from the common previous interactions provided by the content of the history. The identity of Alice is proved by Charlie (and some other meetings) that Bob could verify because he have met Charlie. Of course, a classical strong authentication could be performed for two nodes of the same domain using in our case a zero-knowledge verification but this is not our main aim. 


\section{The Common History Extraction (CHE) protocol} \label{protocol}
In this section we focus on the core of the decision-making process which is called \textit{Common History Exchange} protocol. It is based on cryptographic material (ID-Based Cryptosystem in particular) and some algorithms (e.g., a search algorithm to lookup the presence of an encounter node in a history). In \cite{}, Boneh and Franklin proposed an Identity-Based Encryption (IBE) that permits to us to construct each secret/public key pair endowed for each node in order to build a secure channel and thus to cipher all communication messages. The original motivation for identity-based encryption is to help the deployment of a public key infrastructure. More generally, IBE can simplify systems that manage a large number of public keys. Rather than storing a big database of public keys the system can either derive these public keys from usernames, or simply use the integers {1, ... ,n} as distinct public keys. We also decide to use the Chen-Zhang-Kim's Identity-Based Signature (IBS) scheme as defined in \cite{CZK03} to sign the required elements. Thus, each node will have two pairs of public/secret keys, one pair $(S_{\mathrm{ID}}, Q_{\mathrm{ID}})$ for the cipher operation and one pair $(S^{S}_{\mathrm{ID}}, Q^{S}_{\mathrm{ID}})$ for signature purpose.

Briefly speaking, when two nodes encounter themselves for the first time, they search for possible common nodes they could have met before. To prove an interaction, they have to create together a provable value: this is done simply here by the signature of an agreement or a contract by both parties. The terms of the agreement are called here semantic values and are affected for each element of history. One may derive from these values more information than only the simple intersection of nodes already encountered. This method could be compared to the one used in the {\textquoteleft}Web of trust{\textquoteright} defined by GnuPG \cite{PGP}. However, in our approach, we do not want to prove an identity but just some previous interactions. The identity itself is proved by the use of Identity-Based Cryptography (IBC).

\subsection{Identity Based Cryptosystem} \label{IBE}
Our protocol also inherits some interesting features as its basic cryptographic tools are designed on Elliptic curves.
Identity based cryptography has become a very fashionable topic in the last couple of years. The motivation of this concept, introduced by Shamir in 1984, was to simplify key manage-
ment and avoid the use of digital certificates. The trick was to let a public key be publicly and uniquely derivable from a human-memorizable binary sequence corresponding to an information
non-ambiguously identifying its owner (e-mail address, IP address combined to a user name,
social security number,...) while the associated private keys can only be computed by a trusted
Private Key Generator (PKG) thanks to a master secret $s$. This paradigm allows bypassing the
trust problems that arise in traditional certificate-based public key infrastructures (PKIs). Indeed, since a public key 'is' its owner's identity, it becomes useless to bind them by a digital certificate. Although a PKG's public key still has to be certified, the need of digital certificates is really reduced as reasonably many users may depend on the same PKG.
Since the concept's appearance in 1984, several practical identity based signature schemes
(IBS) have been devised in the late 80's and also after 2001.

On the other hand, finding a practical IBE remained an open challenge until 2001 when Boneh and Franklin proposed to use bilinear maps over algebraic curves to elegantly solve the challenge. After that, these fashionable bilinear maps provided plenty of other applications including various particular kinds of signatures: blind, ring, undeniable, proxy, etc. Along the evolution of public key cryptography from 1976, there has been a graduate evolution tending to a necessity to provide security proofs for asymmetric cryptosystems in the sense that the existence of an attacker against them would imply a probabilistic polynomial time algorithm to solve a hard number theoretic problem. In 1993, motivated by the perspective to achieve provable security for efficient protocols, Bellare and Rogaway introduced the random oracle model (\cite{BellareR93}) that was previously implicitly suggested in \cite{fiat87how} and in which hash functions are used as black box by attackers for whom they are also indistinguishable from perfectly random functions. Although it is well known that security in the random oracle model does not imply security in the real world as shown by several papers exhibiting pathological cases of provably secure schemes for which no secure implementation exists, it still seems to be a good principle to give security proofs 'at least' in the random oracle model when proposing a new asymmetric cryptosystem. In the area of provable security, the last couple of years saw the rise of a new trend consisting of providing tight security reductions for asymmetric cryptosystems (\cite{DBLP:conf/icisc/Pointcheval01}): the security of a cryptographic protocol is said to be tightly related to a hard number theoretic problem if an attacker against the scheme implies an efficient algorithm solving the problem with roughly the same advantage. This led several authors to provide search for new security proofs for systems that were already well known to be secure in the random oracle model or for some of their variants or to devise new schemes that, although appearantly less efficient than existing ones at first sight, provide much better security guarantees for the same security parameters and are then eventually more efficient for a similar desired level of security.

Although concerned with the provable security of identity based signatures, the research
community did not really focus on providing really strong security arguments for the various
IBS proposed in the literature up to now. Indeed, Paterson's IBS still has no formal security
proof while Hess and Cha-Cheon gave proofs under the Diffie-Hellman assumption for their
respective scheme but these proofs were both obtained through Pointcheval and Stern's forking
lemma (\cite{pointcheval96security}) which does not yield tight security reductions as already argued in several
previous papers (\cite{katz-efficiency}).

\subsection{A general view of the CHE protocol}
We denote by "Common History Extraction" a protocol of extraction of common acquaintances contained in the nodes' history. We decide to use the Boneh and Franklin proposition \cite{BF01} to construct the secret/public key pair of each node, to cipher some messages and also to build a secure channel with a weak authentication. We also decide to use the Chen-Zhang-Kim's Identity Based signature scheme as defined in \cite{CZK03} to sign the required elements. Thus, each node will have two pairs of public/secret keys, one pair $(S_{\mathrm{ID}}, Q_{\mathrm{ID}})$ for the cipher operation and one pair $(S^S_{\mathrm{ID}}, Q^S_{\mathrm{ID}})$ for a signature purpose.


In order to establish a trusted interaction, two nodes must prove one to each other that they have a sufficient number of common trusted nodes. Each node builds its history on the base of their previous interactions with its encountered nodes and adds a semantic value describing the nature of the link. For example, if the interaction with another node was a success.

When two stranger nodes meet for the first time, they exchange a part or the full content of the list of their trusted nodes stored in their respective history. Then, they search for possible common trusted nodes that they have met before. According to trust and security policies, the interaction may continue if the number of common nodes is upper a given threshold. Of course, they have to prove one to each other that the interaction really took place. As example, we can consider a node called Alice who has already had a past interaction with Charlie. She could prove it to Bob only if Bob has also had an interaction with Charlie.
Futhermore, if Alice (resp. Bob) trusts Bob (resp. Alice), she (he) adds in her(his) history an entry signed by Bob (resp. Alice). This entry can be checked by anyone having the identity of Alice (resp. Bob) in its history.
We can notice that this kind of interaction is built upon the social principle {\textquoteleft}a friend of my friend is a friend of mine{\textquoteright}. 

So, Alice and Bob have built in a secure channel (created by the IBE scheme) a message denoted $m$ which may be considered as a reciprocal trust. This message is sign respectively by both parties. Alice publishes in her history $(m, Q_{\mathrm{B}}, Q^{S}_{\mathrm{B}}, sign_B(m))$ whereas Bob publishes $(m, Q_{\mathrm{A}}, Q^{S}_{\mathrm{A}}, sign_A(m))$ in his history.

\begin{eqnarray*}
\mbox{A} & \stackrel{\mbox{creation of a secure channel (IBE Scheme)}}{\hbox to 130pt{\leftrightarrowfill}} & \mbox{B} \\
\mbox{A} & \stackrel{\mbox{create a message m={\textquoteleft}ID$_A$ and ID$_B$ trust one to each other{\textquoteright}}}{\hbox to 130pt{\leftrightarrowfill}} & \mbox{B} \\
\mbox{A} & \stackrel{\mbox{A signs m with IBS}}{\hbox to 130pt{\rightarrowfill}} & \mbox{B} \\
\mbox{A} & \stackrel{\mbox{B signs m with IBS}}{\hbox to 130pt{\leftarrowfill}} & \mbox{B} \\
\end{eqnarray*}
In an identical manner, we can consider that when Bob encounters Charlie, they proceed as above
\begin{eqnarray*}
\mbox{B} & \stackrel{\mbox{creation of a secure channel (IBE Scheme)}}{\hbox to 130pt{\leftrightarrowfill}} & \mbox{C} \\
\mbox{B} & \stackrel{\mbox{create a message m'={\textquoteleft}ID$_B$ and ID$_C$ trust one to each other{\textquoteright}}}{\hbox to 130pt{\leftrightarrowfill}} & \mbox{C} \\
\mbox{B} & \stackrel{\mbox{B signs m' with IBS}}{\hbox to 130pt{\rightarrowfill}} & \mbox{C} \\
\mbox{B} & \stackrel{\mbox{C signs m' with IBS}}{\hbox to 130pt{\leftarrowfill}} & \mbox{C} \\
\end{eqnarray*}
When Alice meets Charlie and one to each other want to prove that they have respectively met before Bob, they exchange their public values in their histories and Charlie, first, proves to Alice that Bob trust him using m'. 
\begin{eqnarray*}
\mbox{A} & \stackrel{\mbox{did you meet Bob before ?}}{\hbox to 130pt{\rightarrowfill}} & \mbox{C} \\
\mbox{A} & \stackrel{(sign_B(m'), K_{p_{B}})}{\hbox to 130pt{\leftarrowfill}} & \mbox{C} \\
\mbox{verifies m'} & &  \\
\end{eqnarray*}
The same process is repeated for Alice.
This protocol permits to guarantee the traditional cryptographic properties: authenticity (as Charlie knows the Bob's public key, he could authenticate his signature), integrity is guaranteed by the hash function used in the IBS scheme as in the classical case of a certificate, confidentiality is guaranteed by the secure channel. The secure channel built at the beginning of the exchange in the first step also prevents a man in the middle attack:the reciprocal authentication between Alice and Bob is done by their own trust bond and the secret is only known by the concerned nodes and has been carefully exchanged using a public key cipher method based on IBE.

Notice also that due to the particular structure of the built message, we could easily add some semantic notions in this message and prove the associated keywords used.



\subsection{First studies concerning the size of the history}
The second part of our study concern the required size of the history in the following context: suppose that inside a large group of nodes, you have a particular community that often interacts, the number of common nodes in the history and the size of the history itself must preserve this community (joined together for a semantic purpose as music discussion,...) without drowning it in the large group of nodes. In other words, the chosen parameters must be carefully designed in order to prevent communities from disappearing or from being obliged to always force the interactions ``by the hand''.

\subsubsection{Probabilistic approach}

We consider here that the size of the history is $k$ and depends on the total number $n'$ of nodes for a given imprinting station. We suppose that the size of the small community is $n$. We then want to estimate the required number $p$ of common nodes in the history to permit the access to some services. We suppose in this subsection that the nodes meetings are random and does not depend on some laws of proximity. A more sophisticate random graph will be studied in the simulation paragraph.

When two nodes $A$ and $B$, belonging to the same small community, meet each other and want to know the number of common nodes they have in their own histories. They could exchange some services if they have at least $p$ common nodes in their history of size $k$. We want that the probability that the previous event happens is very high otherwise, this probability must be low.

Those two probabilities derive in fact from the same computation that is the following in the case of a group of size $n$: we first compute the probability that they have exactly $i$ common nodes in their histories. This probability is given by : $\frac{\binom{n}{i} \binom{n-i}{k-i} \binom{n-k}{k-i}}{(\binom{n}{k})^2}$. (This corresponds to the way to take $i$ values inside $n$ possibles and then the $k-i$ staying values inside the $n-i$ nodes and last the $k-i$ values inside those that do not belong to the history i.e. $n-k$.) This formula is true if $k-p \leq n-k$ i.e. $p \in \{\max(2k-n,0),k \}$ (the size of the history must be upper bounded by $n/2$).

We then deduce the probability that $A$ and $B$ have at most $p$ common knowledges (excluding $p$): $\frac{1}{(\binom{n}{k})^2} \cdot \sum_{i=0}^{p-1} \binom{n}{i} \cdot \binom{n-i}{k-i} \cdot \binom{n-k}{k-i}$.

And then, the probability $P$ they have at least $p$ common knowledges is given by:
\begin{eqnarray}
P=Pr(A \cap B \geq p)& = & 1-\frac{1}{(\binom{n}{k})^2} \cdot \sum_{i=0}^{p-1} \binom{n}{i} \cdot \binom{n-i}{k-i} \cdot \binom{n-k}{k-i} \label{eq}
\end{eqnarray}
This probability stays true in the case where the nodes only belongs to the same domain and where $n$ is replaced by $n'$. We call this probability $P'$.

We have computed those probabilities considering that we want that $P'$ must be high and $P$ low. We obtain the following results according the $p$ and $k$ values.

\begin{center}
\begin{tabular}{|c|c|c|c|c|c|}
\hline
$n'$ & $n$ & $k$ & $p$ &  P & P' \\
\hline
30 & 100 & 6 & 3 & 0,2 \% & 7,6 \%\\
\hline
30 & 100 & 8 & 3 & 1,6 \% & 35,5 \%\\
\hline
30 & 100 & 10 & 3 & 6 \% & 74,9 \%\\
\hline
30 & 100 & 12 & 3 & 15,5 \% & 96,2 \%\\
\hline
30 & 100 & 6 & 4 & 0 \% & 0,7 \%\\
\hline
30 & 100 & 8 & 4 & 0,1 \% & 10,3 \%\\
\hline
30 & 100 & 10 & 4 & 0,8 \% & 43,96 \%\\
\hline
30 & 100 & 12 & 4 & 3,56 \% & 83,3 \%\\
\hline
30 & 100 & 6 & 6 & 0 \% & 0 \%\\
\hline
30 & 100 & 8 & 6 & 0 \% & 0,1 \%\\
\hline
30 & 100 & 10 & 6 & 0 \% & 4 \%\\
\hline
30 & 100 & 12 & 6 & 0 \% & 29,6 \%\\
\hline
\end{tabular}
\end{center}

\subsubsection{Simulation approach}

The KAA model is dedicated to smart devices, therefore such devices are belonging to a person and so resulting interaction graph is a social graph.  Social graphs have been studied for a long time, first by sociologists and more recently by mathematicians. 

The first property of social graph is the {\it small world} effect. This property means that even in social graph strongly geographical (so with insular part or social barriers) there exists short connecting path. More recently, some works emphasize recurrent clustering organization which can also affect the way social graph should be studied. The last property, which is very important for simulation, is  the skewed degree distribution. 

In order to study the $p$ parameter of the KAA trust model we use random graph with skewed distribution. 
The sequence of degree is obtained through an  exponential and continuous power-law distribution generator.

The aim of our simulation is to provide basic idea to verify the correct choice of the $p$ parameter for a given community. The goal is to choose the right $p$ parameter that give large probability of spontaneous interaction between nodes of the community and low probability of interaction between a node not belonging to the same community. This empirical approach need the knowledge of the community, in term of degree distribution. Most of community specification are arbitrary. This is a first step to automatic - or semi automatic, configuration a KAA model for a specific community.

Let us suppose a community denoted $C$ of $30$ nodes interacting inside a social group $G(V,E)$ of $|V|=100$ nodes, including the $C$ community. We suppose that interactions are more frequent between nodes of $C$ than between a node of $C$ and a node of $G\backslash C$. Therefore, $G$ also constitutes a social group and has same general properties. In our simulations, we define a community with $4$ parameters: $s$ the size of the community, $d_{min}$ and $d_{max}$ corresponding to the range of possible degrees of nodes, and $\alpha$ the exponent of the power-law distribution function. Here the parameters both of $C$ and $G$:

\begin{center}
	\begin{tabular}{|c|c|c|c|c|}
	\hline
	 & Nodes & $d_{min}$ & $d_{min}$ & $\alpha$\\
	\hline
	$C$ & 30 & 6 & 12 & $2.4$\\
	\hline
	$G$ & 100 & 5 & 10 & $2.4$ \\
	\hline		
	\end{tabular}
\end{center}

In order to evaluate the cohesion of the community, we define a notion of distance between nodes of the community. We say that $y \in C$ is in the neighborhood of $x \in C$ if there exists an edge between $x$ and $y$ in the graph of $C$ (a trust relation have been already established). We call this first ring neighborhood $V_1$ and generalize this notion as follow: 

\[ V_1(x) = \{y | \exists x, (x,y) \in E \} \]

\[ V_i(x) = \{y | \exists z_1,\cdots z_p, (x,z_i) \in E \mbox{ and } (z_i,y) \in E \} \bigcup V_{i-1}(x)\]

\begin{center}
\begin{figure}
\begin{center}
\includegraphics[height=7cm]{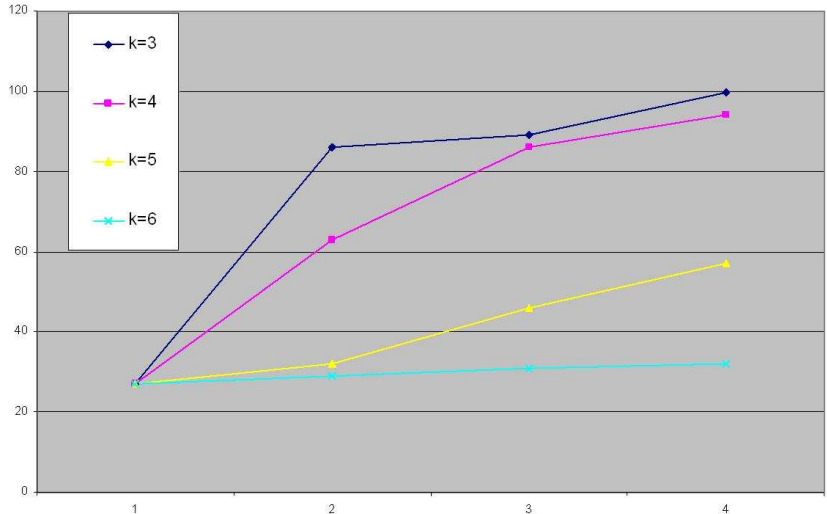}
\end{center}
\caption{Impact of the $p$ parameter of the size of the neighborhood.}
\end{figure}
\end{center}

Therefore, with the same $p$ parameter, we do not wont such properties to be easily transmitted to nodes outside of $C$. In our experiments, usually $1$ or $2$ nodes from $G\backslash C$ 	are at distance less $4$ to a node of $C$ with $p>3$.   
Our model seems to be rather resistant to automatic incorporation of nodes out of the initial community.


\section{Reputation Model} \label{reputation}
\subsection{Semantic of the history content}

First, we observe that each interaction is asymmetric with a sender terminal and a receiver terminal that receives the asked service. Even in a collaborative action, it is possible to isolate each action in an asymmetric way.
If we consider this asymmetric notion in the creation of an element of the history, we could extract from this fact a semantic value that will lead our trust policy.

We also assume here that there exists no protocol that permits to guarantee a simultaneous exchange for the generation of a history element in each terminal and that there is no mechanism that guarantees that the generation of a history element will really take place. Moreover, as soon as a terminal refuses this generation (because it's against its own policy or because it is a cheater node), no history element will be generated.

If we take into account all the previous remarks, we then study the different policies that configure a node reaction. Clearly, we have to force the history elements generation in order to enrich the global network using incitations. We also take into account the human reciprocal behavior in such context as proved by the reactions observed in experimental economic games or in the peer-to-peer networks where the first aim is the existence of the network. \\

\subsubsection{The generation mechanism for the history elements.} \label{algo}

The first step consists of deciding at which moment the generation of a history element will begin. Of course, this generation could not happen before the exchange of services due to the fact that a cheater node could build a history without providing services and in this case, the model would not be anymore a trust framework. So, the generation of the history elements must be performed after the service realization. In this part, we do not integrate a semantic value for all possible services. We just consider here a successful interaction where the service is really provided. \\

It seems logical that the node that first gives a history element is the node that receives the service from another. Thus, a terminal with a rich history could interact more easily with others.\\
 
Then, our algorithm could be written as follows: 
\begin{itemize} 
\item \textbf{First step (1):} suppose that the node A (the receiver) asks the node B (the provider) for a particular service.
\item \textbf{Second step (2):} The node B could provide the service to the node A. In this case, the message formed to be signed by the two parties in presence will be $m=$'B provided a service to A'. In the case where the service is not furnished by B, the algorithm stops here. We call this property $sp_A(B)$='service property'.
\item \textbf{Third step (3):} The node A could enrich the history of B by signing and sending to B the message $m$. In this case, we say that A is a trustor and we call this property the $tp_B(A)$='trustor property'. If the node A does not provide the history element, the algorithm stops due to the 'non trustor property'.
\item \textbf{Fourth step (4):} The node B has previously received from the node A the $tp_B(A)$ and it could be reciprocal: it could provide to the node A a history element signing and sending the message $m$. We call this property $rp_A(B)=$'reciprocal property'. Of course, the node B is not obliged to provide this history element and could verify the 'non reciprocal property'.
\end{itemize} 

\bigskip

This algorithm is illustrated on the following figure:
 
\begin{eqnarray*}
\mbox{A} & \stackrel{\mbox{asks B for a service}}{\hbox to 130pt{\rightarrowfill}} & \mbox{B $\; \;$ (1)} \\
\mbox{A} & \stackrel{\mbox{B provided the service}}{\hbox to 130pt{\leftarrowfill}} & \mbox{B $\; \;$ (2)} \\
\mbox{A} & \stackrel{\mbox{A signs $m$=B provided a service to A}}{\hbox to 130pt{\rightarrowfill}} & \mbox{B $\; \;$ (3)} \\
\mbox{A} & \stackrel{\mbox{B signs $m$=B provided a service to A}}{\hbox to 130pt{\leftarrowfill}} & \mbox{B $\; \;$ (4)} \\
\end{eqnarray*}

In step (3), B keeps $sign_A(m)$ while A keeps $sign_B(m)$ in step (4).

Once this algorithm is implemented, we need to add the following remarks on the cryptographic protocol used:
\begin{itemize}
\item the node A is not obliged to provide the $tp_B(A)$ property.
\item the same remark holds for the $rp_A(B)$ property.
\item The cryptographic protocol used guarantees that the history element could only be used by the node that receives this element (nodes A and B here). 
\item The only sanction that a node could take if it is not satisfied by an action (the non $sp$, non $tp$ or non $rp$ properties) is to blacklist the non trusted node. This action implies that the non trusted node could no more interact with the deceived node. (We suppose here that a blacklist operation will only take place after a threshold of non trusted interactions).   
\item The KAA framework supposes that there is no recommendation model and that no central authority could act for additional sanctions.
\end{itemize}

\subsubsection{Semantic notions.}

For an interaction such the one previously described in section \ref{algo} between two nodes A (receiver) and B (provider), some semantic notions could be released for each node.\\

The history element received by B contains the following semantic notions:
\begin{itemize}
        \item B provided a service to A: $sp_A(B)$ property.
        \item A was a trustor for B: $tp_B(A)$ property.
\end{itemize}

\bigskip

The history element received by A contains the following semantic notions:
\begin{itemize}
        \item B provided a service to A: $sp_A(B)$ property.
        \item A was a trustor for B: $tp_B(A)$ property.
        \item B was reciprocal for A: $rp_A(B)$ property.
\end{itemize}

\bigskip

Of course, A is incited to be a trustor because this is the only way to enrich its own history. But, a \textit{free rider} behavior could also be attended. However, B has not a real interest to be reciprocal except if it takes into account the general enrichment of the network and of its personal reputation with A first and with the others if a reputation mechanism is set up. \\

Notice first that all the two last properties are proved by the cryptographic algorithm used in our model, even if we could not prove an interaction with no exchange of history elements. The two elements of history built on the two exchanges will be, from now, called trustor proof for the first one and reciprocal proof for the second one. 

When two terminals enter in an interactive session, they will inverse the two roles during the session and will possess the two elements of history.

\subsection{Reputation model} \label{reput}

The KAA framework objects the idea of a recommendation mechanism, the trust being considered in our model in a non transitive way. Indeed, the main principle of the KAA framework is to base our policy only upon cryptographic provable elements. Consequently, a node could not receive a recommendation for a node that it has never met before because there is no simple and local means to prove the semantic of this recommendation.

However, the KAA framework permits the management of a reputation mechanism, locally computed, as soon as it will be based upon provable elements. First, as we noticed before, the interactions could happen between nodes of the same community or between nodes from different communities and could also be proved. 

The computation of the reputation value of the node A for the node B could be performed using the following informations computed at two levels:

\begin{itemize}
        \item \textbf{Direct} The node A keeps the memory of past interactions between A and B, using the semantic values generated from the history elements. This information could be proved by a third party who has already met B. We suppose that history elements are marked by a time stamp else the trust bond is definitely acquired. The history element will only indicate the quality of the sequence and the number of past interactions with a single node ('trustor proof', 'reciprocal proof' or both in case of an exchange). This number could not be transfered because it is not provable. All those values are cryptographically proved and this model avoids some nodes coalitions to destroy the reputation of one node. The node A that is convinced by those elements could derived from those values a reputation value for the node B.
        \item \textbf{Indirect} When A interacts with a new node C, they exchange whole or parts of their histories. If they have B as a common knowledge, A could receive from C the semantic of the past interactions between B and C ('trustor proof', 'reciprocal proof' or the both). The node A could take into account those values to update its own reputation value for B. Of course, there is no information about the quantity of previous interactions that C has built with B. This bonus of reputation could be taken into account only once for each correspondent other than B and that presents B in its own history.             
\end{itemize}

We could then notice that the trust indirect policy does not violate the bases of the KAA framework: we only use trust values that have been cryptographically proved. It is impossible to create by its own an entry in the reputation table for a particular node never met before. Moreover, this reputation model is an incitation to be reciprocal and let a proof of good behavior in the histories of nodes met before. Thus, the nodes could hope to have a good reputation for a lot of system nodes.

%
\subsection{Some elements for a trust policy} \label{policy}

As previously described, the aim of the KAA project is to build a framework for many trust management systems rather than a specific one. With the two basic mechanisms of the KAA model - namely common history and reputation - one may build a wide range of trust policy. Let us sum up first all parameters that can be adapted. Adaptation can be static, depending of the social pattern the user of the smart device is belonging to or dynamic depending of the context of use. 

The main parameter of the KAA model is the size of the common history a node requires to accept an interaction. Note that, there exists an asymmetry in interactions and the size of the common knowledge required to be receiver or to be provider not needs to be the same. Also the fact that the corresponding node belongs or does not belong to the same community (security domain) may also have an impact of that. For intra domain relationship, the size of the common history required is clearly related to the size of the community. 

We have computed the corresponding probability of success in a such case and, inspired from the birthday paradox, have observed that for a given group of size $n$, if the size $k$ of the history is $n/ \ln(n)$ and the threshold number $p$ of common knowledges is about $\sqrt{n/ \ln(n)}$ then the probability of success (here to create a trust link) is greater than 50$\%$. So, for example, if $n=100$, $k=22$ and $p=5$, the success probability is about $56,6 \%$ (for the same parameters and $p=3$, this probability reachs $92 \%$). In this case, we could see that the size $k$ of the history is reasonable and could be easily carried by each node and that the number of verifications to perform given by the $p$ value is also not excessive. 
 
With those requirements, we need also to tell that the history works on the principle of a crush FIFO: the first input of the history corresponds with the first output when the history is full.\\

So the node A could use the previous remarks to build a trust policy based upon the following notions:

\begin{itemize}
\item \textbf{Direct (for a node):} number of meetings with a node, number of trustor proofs, number of reciprocal proofs, number of services refused, last encounter common knowledge size, date of the last blacklist enter.
\item \textbf{Indirect (for a node):} number of encountered nodes carrying a trustor proof, number of encountered nodes carrying a reciprocal proof,  number of encountered nodes carrying both  trustor/reciprocal proof.
\end{itemize}


\subsection{Social pattern examples} \label{exple}
 
Our trust policy depends on the type of exchange as defined in section \ref{social} and is based upon the social patterns. To reach this aim, we have defined two particular modes of operation: a closed mode where, as in the human society (the family for example), the trust pre-exists and in our model, the cryptographic verifications are not required and an open mode, where the trust bond need to be established using the CKE protocol. This mode could be used for example in an organization pattern as a company. According the size of the company and the incurred risk, trust policy will adjust the parameters previously defined and the required number of common knowledges. 

\begin{itemize}
\item \textbf{Family:} a community with a strong social distance and a strong degree of structure works in the closed mode.
\item \textbf{Network:} a community with a strong social distance and a weak degree of structure works in the open mode using a weak authentication for the nodes meetings.
\item \textbf{Market:}
To implement a Market-like community, a money (either virtual or not) is required. Therefore, it is possible to add a semantic value to trustor proof and reciprocal proof based on the notion of market. If you add the value of the transaction in both proofs, you can achieve this goal. The trustor proof and reciprocal proof can be automatically generated when the transaction took place (security of this transaction is out of the scope of this paper).
\item \textbf{Organization:} In an organization pattern as a company, the closed mode is used with a strong authentication delivered by the imprinting station through classical certification. We could also integrate the hierarchical parameter in our model: for example, if someone wants to talk to a leader, the required number of common history elements need to be larger than if he wants to talk to a person at the same level than him.
\end{itemize}

Our model is also an inter domain framework: for example, a closed community could act being altruist. This community always provides the asked services but it is never reciprocal to preserve itself: it does not want to increase the number of the history elements of the outer nodes in order to prevent those nodes from entering inside the community.

\section{Further Works} \label{further}
\subsection{Trust policies}
From a trust decision cycle point of view, \cite{edmst} shows that trust could be seen as a comsumer/provider one-way relationship because it defines a set of distinct properties for each entity involved in the exchange. Each entity could become a consumer, a provider or the both at the same time. Then, the trust relationship between a consumer and a provider translates a set of actions or just a single one-to-one action \cite{grandison00survey}. Each action possesses its own intention and its own objective. The intention defines the activity of the consumer on the provider and can take three states: positive, negative or suspect according to the match between the intention and the trust policy. The objective notion describes the activity type of the entity. Thus, a consumer shows its positive intention if its objective is only, for example, to print a file on the provider's printer but it shows its negative intention if its objective is to inject a virus code on the consumer's laptop.

The trust relationship of a provider-toward-consumer's system is a process which modifies the initial state of the provider's system into a resultant state (consequence of the action) necessarily conform with the rules established by the provider. Those rules describe the trust policy of the entity, seen as a provider or a consumer. For \cite{grandison00survey},  those rules are built for each entity from an opinion notion, a judgment concept and some recommendation mechanisms. The opinion notion as described in \cite{josang} combines belief, unbelief and ignorance values for each action or for each entity in order to construct a recommendation. This aspect is very difficult to measure. To see those three values as some boolean ones taking 'True', 'False' or 'I don't know' is the easiest representation. Moreover, other factors influence the opinion that the provider builds upon the consumer as the past experience sharing (the history in our case) or the gain or the loss inherently carried by the action. 
\cite{Carbone}, \cite{edmst} show that the decision cycle must lead to a decision. This one is a boolean value that exists if and only if each entity takes the same decision. However, if only a single entity takes an opposite decision, the exchange is locked. So, to prevent a such situation from happening, a negotiator model could be implemented. This mechanism ensures that all the exchange necessary conditions are collected and it provides one or many trade-offs. Those trade-off mechanisms are going to take into account some new parameters as the history of previous interactions, the number of such interactions and an evaluation of those meetings.

The decision life cycle regulation becomes the difficulty of a such model and we think that the trust policies could be compared with a contract negotiated between both a consumer and a provider. With this intention, a risk management model seems to be the most convenient representation to express decision management trust policies based upon decisions.

Our main goal is to define trust policies in order to express the notion of risk management through the effects on the system state that a particular activity could imply. The decision making process will be influenced by several criteria such as the experience that constitutes the inheritance of the entity knowledge. The decision life cycle is represented by three fundamental notions that guarantee the decision making:
\begin{itemize}
\item The environment analyze that leads to the knowledge of the exchange context.
\item The knowledge that represents the experiment learned near the entities previously met.
\item The trust policies also called ``dogma'' that define the rules set previously learned to regulate the trust. As proposed in \cite{Carbone} and \cite{edmst}, the risk policies take into account the cost of the resources furnished by the provider and the exposition of this particular resources.   
\end{itemize}

Risk management applied into our trust model must respect many criteria described below:
\begin{itemize}
\item First, exchange depends directly and always on its context; this means that entity is able to discover its current context, then take it into account in its analysis and still in its decision process. We believe that adaptive trust protocol is an essential functionality.
\item Second as described before, we must distinguish two types of entities exchange according intention and objective: the consumer type entity and second the provider type entity. So an entity can be the both.
\item Third, an entity must be able to implement a relation even if no entity of its environment is known by it. The decision making mechanism must solve this aspect and exceed the security policy that leads most of the time on the decision of a blocked exchange.
\item Fourth, the notion of acquired experiment (the history in our case) will distinguish four conditions and four trust policies according to the entity performs its first exchange, performs its first exchange with a given entity (the exchange is authorized because the number of previous meetings with some common nodes is upper a given threshold or the exchange is not authorized because the condition of the trust policy is not met), has previously exchanged with this entity. For example, in case of a first exchange, our decision policy must be able to take into account a loss of experiment. The notion of share experiment will be a main aspect of our trust policy. 
\item Fifth, two entities that decide to exchange must be able to modify their personal judgments according some external informations. Those informations depend on the context and can be increasingly personal for an entity or a community. In such a context, we define the notion of the semantic of the exchange that will enrich in a finer way the decision making process. 
\end{itemize}

We are able now to represent the trust management process implemented in the decision making process:
\begin{itemize}
\item updates of the trust criteria.
\item history management.
\item control management using a pro-active mechanism.
\item decision making process with an adapted language.
\item negotiation of the relation conditions.
\item relation conditions making process.
\item detection of the criteria necessary for the trust negotiation.
\end{itemize}

Our aim is also to introduce a policy language and to use it in our trust structure to define a decisional semantic in our model. 

\subsection{Formal proof}
The main difference between the informal notion of history-based
explored in the social sciences and formality needed for computing is
that in the end, our model has to be operational. Also, this model has
to be twofold: first, to include a notion of protocol to describe the
exact behavior of systems, which is fundamental when security is
concerned and second,  to allow nodes' interactions to feedback to the
security mechanisms and influence future policies. Also, we propose a
formal model for history-based trust management systems in ambient
networks providing these two aims. In this model, a node will be then
specified by a pair, a policy $\alpha$ and and a protocol $P$
interacting in the following way. The policy $\alpha$ informs the
protocol  $P$ about what actions are allowed at any moment and works
on the basis of the history of past interactions. Dually, $P$
interacts with the network of other ambients and doing so, it produces
the observations gathered in $\alpha$. The policy will be specified
with a decidable logic while the protocol will be modeled with a
process calculus {à la\/} $\pi$-calculus.

\section*{Conclusion} \label{conclusion}
\addcontentsline{toc}{section}{Conclusion}
In the context of \textit{Ambient network}, mobiles will often become disconnected from their home networks and will have to handle unforeseen circumstances. The mobile needs to carry self-contained informations and methods to be able to make fully autonomous security decisions.
In this paper, we present a general framework for managing trust in a fully distributed environment. We propose to record the resulting data of the interaction of mobiles in a \textit{History}. These data are made secure by cryptography tools. Then when
two nodes encounter themselves for the first time, an interaction could take place if the
number of common trusted nodes is greater than a specified threshold. We consider that our approach is relevant by the fact that history is secure and it is not-transferable except against the clone attack. The \textit{Common History Extraction} is a protocol designed both to generate elements of history and also to verify their authenticity: it is a full part of the decision-making process. Based on the elliptic curves, our protocol may be embedded in small devices such as PDA or smart phones.

We have also shown how these data should be combined to build a flexible trust management depending on the context: semantic values may be added to elements of history and that allows our protocol to be in adequacy with many real environments. Mobiles will have the capability to interact according to their trust policies since they will share the same cryptographic material although they belong to different domains.To sum up, the key feature of our framework is that trust is based on a local reputation system and is build on the own-experience of the mobiles.

\section*{Acknowledgment}
\addcontentsline{toc}{section}{Acknowledgment}
This work is done in the \textit{Knowledge Authentication for Ambient} (KAA) project supported by the ACI-SI program of the French ministry of research. The professor John Mullins from the Ecole Polytechnique de Montr\'eal joins the research while visiting the CITI Lab and the INRIA ARES project during his sabbatical.

\bibliographystyle{plain}
\bibliography{secu_sub}


\appendix

\section{Bilinear pairings}\label{pairing}
Here we briefly give properties of a cryptographic bilinear map which is a modified Weil pairing \cite{BF01}.
A cryptographic bilinear pairing is defined as
$e : G_1 \times G_1 \to G_2$ where $G_1$ is an additive cyclic group of prime order $q$,
$G_2$ is a multiplicative cyclic group of the same order and $P$ is an arbitrary generator of $G_1$.
An admissible bilinear pairing has the following properties :
\begin{itemize}
\item\textsf{Bilinear}: $e(aR,bS) = e(R,S)^{ab}\quad \forall R,S \in G_1$ and $a,b \in \mathbb{Z}^*_q$.
This can be restated as
$\forall R,S,T \in G_1$, $e(R+S,T) = e(R,T)e(S,T)$ and $e(R,S+T) = e(R,S)e(R,T)$.\\
\item\textsf{Non-degenerate}:
There exists $R,S\in G_1$ such that $e(R,S) \neq I_{G_2}$ where $I_{G_2}$ denotes the identity element
of the group $G_2$.\\
\item\textsf{Computable}: There exists an efficient algorithm to compute $e(R,S)\quad \forall R,S \in G_1$.
\end{itemize}

\end{document}